\documentclass[10pt,conference]{IEEEtran}

\usepackage{cite}
\usepackage{amsmath,amssymb,amsfonts}
\usepackage{algorithmic}
\usepackage{graphicx}
\usepackage{textcomp}
\usepackage{xcolor}
\usepackage{booktabs}       % Professional tables
\usepackage{listings}        % Code listings
\usepackage[T1]{fontenc}
\usepackage{lmodern}        % Latin Modern fonts (monospace support)
\usepackage{url}
\usepackage{hyperref}
\usepackage{balance}         % Balance last page columns
\usepackage{multirow}
\usepackage{subcaption}      % Sub-figures
\usepackage{tikz}            % Architecture diagram
\usetikzlibrary{positioning,arrows.meta,fit,backgrounds,calc}

\def\BibTeX{{\rm B\kern-.05em{\sc i\kern-.025em b}\kern-.08em
    T\kern-.1667em\lower.7ex\hbox{E}\kern-.125emX}}

\begin{document}

% ══════════════════════════════════════════════════════════════════════
% TITLE & AUTHORS
% ══════════════════════════════════════════════════════════════════════

\title{SAGAI-MID: A Generative AI-Driven Middleware\\for Dynamic Runtime Interoperability}

\author{
  \IEEEauthorblockN{Oliver Aleksander Larsen}
  \IEEEauthorblockA{
    \textit{SDU Software Engineering}\\
    University of Southern Denmark\\
    Odense, Denmark\\
    olar@mmmi.sdu.dk
  }
  \and
  \IEEEauthorblockN{Mahyar T. Moghaddam}
  \IEEEauthorblockA{
    \textit{SDU Software Engineering}\\
    University of Southern Denmark\\
    Odense, Denmark\\
    mtmo@mmmi.sdu.dk
  }
}

\maketitle
\footnotetext{\textcopyright~2026 IEEE. Personal use of this material is permitted. Permission from IEEE must be obtained for all other uses, in any current or future media, including reprinting/republishing this material for advertising or promotional purposes, creating new collective works, for resale or redistribution to servers or lists, or reuse of any copyrighted component of this work in other works. Accepted at SAGAI 2026, co-located with IEEE ICSA 2026.}

% ══════════════════════════════════════════════════════════════════════
% ABSTRACT
% ══════════════════════════════════════════════════════════════════════

\begin{abstract}
Modern distributed systems integrate heterogeneous services, REST APIs with different schema versions, GraphQL endpoints, and IoT devices with proprietary payloads that suffer from persistent schema mismatches.
Traditional static adapters require manual coding for every schema pair and cannot handle novel combinations at runtime.
We present SAGAI-MID, a FastAPI-based middleware that uses large language models (LLMs) to dynamically detect and resolve schema mismatches at runtime. The system employs a five-layer pipeline: hybrid detection
(structural diff plus LLM semantic analysis), dual resolution strategies (per-request LLM transformation and LLM-generated reusable adapter code), and a three-tier safeguard stack (validation, ensemble voting, rule-based fallback). We frame the architecture through Bass~et~al.'s interoperability tactics, transforming them from design-time artifacts into runtime capabilities.
We evaluate SAGAI-MID on 10~interoperability scenarios spanning REST version migration, IoT-to-analytics bridging, and GraphQL protocol conversion across six LLMs from two providers. The best-performing
configuration achieves 0.90~pass@1 accuracy. The CODEGEN strategy consistently outperforms DIRECT (\mbox{0.83} vs \mbox{0.77} mean~pass@1), while cost varies by over 30$\times$ across models with no proportional
accuracy gain; the most accurate model is also the cheapest. We discuss implications for software architects adopting LLMs as runtime architectural components.
\end{abstract}


% ══════════════════════════════════════════════════════════════════════
% I. INTRODUCTION
% ══════════════════════════════════════════════════════════════════════

\section{Introduction}
\label{sec:introduction}

Modern distributed systems increasingly integrate heterogeneous
services: REST APIs with diverging schema versions, GraphQL endpoints,
and IoT/MQTT devices with proprietary payloads. Schema mismatches---differences
in field naming conventions (e.g., \texttt{camelCase} vs.\ \texttt{snake\_case}),
data types (ISO~8601 timestamps vs.\ Unix epochs), measurement units
(Celsius vs.\ Fahrenheit), and nesting structures (flat vs.\ nested
JSON)---are a persistent interoperability challenge. Lercher~et~al.\
\cite{lercher2024} find that microservice API evolution is a major
maintenance burden, with loose coupling providing no immediate feedback
on breaking changes.

Traditional approaches rely on static adapters, XSLT transforms, or
configuration-based mapping engines. These solutions require manual
coding for every new schema pair and cannot handle novel combinations at
runtime. Bass~et~al.'s interoperability tactics~\cite{bass2021} provide
a well-established vocabulary for such adaptation, but remain
traditionally bound to design-time implementation.

Recent work demonstrates that LLMs can generate schema mappings and
transformation code with promising accuracy. Falcao~et~al.\
\cite{falcao2026} show that LLM-based DIRECT and CODEGEN strategies
achieve up to 93\%~pass@1 on interoperability tasks, while
Parciak~et~al.\ \cite{parciak2024} find that majority voting
reduces LLM hallucination in schema matching. However, no prior work
{\em 1)}~frames LLM middleware through established architectural tactics,
{\em 2)}~treats safeguards against LLM non-determinism as first-class
architectural concerns, {\em 3)}~evaluates across multiple protocols
(REST, GraphQL, IoT), or {\em 4)}~provides a pluggable middleware
integration pattern.

This paper presents SAGAI-MID, a generative AI-driven middleware for
dynamic runtime interoperability. Our contributions are:
\begin{enumerate}
  \item A five-layer middleware architecture that connects LLM-based
    schema resolution to Bass~et~al.'s interoperability tactics,
    transforming them from design-time artifacts into runtime capabilities.
  \item A hybrid detection approach combining deterministic structural
    analysis with LLM-powered semantic analysis for naming and unit
    mismatches.
  \item Two resolution strategies, DIRECT (per-request LLM
    transformation) and CODEGEN (LLM-generated reusable Python
    adapters), with hash-keyed caching for deterministic replay.
  \item A three-tier safeguard stack (Pydantic validation, ensemble
    voting, rule-based fallback) addressing LLM non-determinism.
  \item Empirical evaluation on 10~diverse interoperability scenarios
    across six~LLMs from two providers, measuring correctness,
    performance, cost, and safeguard impact.
\end{enumerate}

\noindent The implementation and all evaluation artifacts are available as open-source software.\footnote{\url{https://github.com/Oliver1703dk/sagai-mid}}

The remainder of this paper is structured as follows.
Section~\ref{sec:related} discusses related work.
Section~\ref{sec:architecture} presents the SAGAI-MID architecture.
Section~\ref{sec:implementation} describes the implementation.
Section~\ref{sec:evaluation} reports our evaluation results.
Section~\ref{sec:discussion} discusses findings and limitations.
Section~\ref{sec:conclusion} concludes with future directions.

% ══════════════════════════════════════════════════════════════════════
% II. BACKGROUND & RELATED WORK
% ══════════════════════════════════════════════════════════════════════

\section{Background \& Related Work}
\label{sec:related}

\subsection{Interoperability Tactics}
\label{sec:related:tactics}

Bass, Clements, and Kazman define interoperability as the degree to
which two or more systems can usefully exchange meaningful
information~\cite{bass2021}. Their interoperability tactics provide a
vocabulary for architectural decisions: \emph{Discover}, \emph{Tailor
Interface}, \emph{Convert Data}, \emph{Manage Resources}, and
\emph{Orchestrate}. Complementary frameworks include Tolk and
Muguira's Levels of Conceptual Interoperability Model
(LCIM)~\cite{tolk2003}, which distinguishes seven levels from
technical to conceptual interoperability, and Hohpe and Woolf's
enterprise integration patterns~\cite{hohpe2003}, which catalog
messaging-based mediation strategies. The ISO/IEC~25010 quality model
identifies interoperability as a key product quality
characteristic~\cite{iso25010}.

These tactics are traditionally implemented as design-time artifacts:
coded once, tested, deployed, and brittle to schema evolution. Our key
insight is that LLMs can transform these tactics into \emph{runtime
capabilities}, generating the adaptation logic on demand for novel
schema combinations without manual coding or redeployment.

\subsection{LLM-Based Schema Matching}
\label{sec:related:schema}

Recent work applies LLMs to schema matching and data transformation.
Parciak~et~al.\ \cite{parciak2024} evaluate four prompting strategies
for schema matching with GPT-4, achieving F1\,=\,0.58, and find that
majority voting reduces hallucination from 24\% to 8\%, directly
inspiring our ensemble safeguard (Tier~2). Seedat and van~der~Schaar~\cite{seedat2024} propose Matchmaker, a self-improving LLM program
for schema matching via zero-shot self-improvement. Narayan~et~al.\
\cite{narayan2022} demonstrate that foundation models can perform
data integration tasks (schema matching, entity resolution, data
imputation) competitively with task-specific approaches. Peeters
et~al.~\cite{peeters2025} show that LLMs achieve state-of-the-art
entity matching accuracy with zero-shot prompting.

These works focus on \emph{offline} matching for dataset integration.
We apply similar techniques at \emph{runtime} within a middleware
context, requiring both caching (to amortize LLM costs) and a
deterministic fallback path (to guarantee forward progress).

\subsection{LLM Middleware for Interoperability}
\label{sec:related:middleware}

Most directly related to our work, Falcao~et~al.\ \cite{falcao2026}
evaluate LLM-based interoperability using DIRECT (per-request LLM
transformation) and CODEGEN (LLM-generated adapter code) strategies
across 13~open-source LLMs, achieving up to 93\%~pass@1. They show
that CODEGEN outperforms DIRECT on average and that model selection
matters more than prompting strategy. Their evaluation framework and
strategy definitions form the foundation that our resolution engine
builds upon.

Guran~et~al.\ \cite{guran2024} propose a two-tier LLM middleware
architecture with two deployment patterns: \emph{LLM-as-Service}
(the LLM is the primary service endpoint) and \emph{LLM-as-Gateway}
(the LLM mediates between client and backend). SAGAI-MID falls into
the LLM-as-Gateway pattern, where the middleware intercepts traffic
and applies LLM-driven transformations transparently. However,
Guran~et~al.\ do not address schema matching specifically, nor do
they provide empirical evaluation.

Esposito~et~al.\ \cite{esposito2026} survey generative AI for software
architecture, identifying runtime interoperability as an open direction
with ``high potential but limited empirical evidence.'' Self-adaptive
architectures using the MAPE-K loop~\cite{kephart2003} have explored
runtime adaptation through monitor-analyze-plan-execute cycles, but
not LLM-driven schema transformation specifically.

SAGAI-MID builds on Falcao~et~al.'s strategies but makes three
novel contributions: {\em 1)}~embedding the strategies in a five-layer
middleware architecture, {\em 2)}~adding safeguards against LLM
non-determinism as a first-class concern, and {\em 3)}~connecting to
established interoperability tactics. We additionally evaluate across
multiple protocols (REST, GraphQL, IoT) and commercial LLM providers,
whereas prior work evaluates one protocol with open-source models.

% ══════════════════════════════════════════════════════════════════════
% III. SAGAI-MID ARCHITECTURE
% ══════════════════════════════════════════════════════════════════════

\section{SAGAI-MID Architecture}
\label{sec:architecture}

\begin{figure}[t]
  \centering
  \begin{tikzpicture}[
    node distance=0.6cm,
    layer/.style={draw, rounded corners, minimum width=5.2cm,
      minimum height=0.65cm, font=\scriptsize, align=center},
    llm/.style={draw, dashed, rounded corners, minimum width=1.2cm,
      minimum height=0.45cm, font=\scriptsize, fill=gray!10},
    arrow/.style={-{Stealth[length=2mm]}, thick},
  ]
    % Layers
    \node[layer, fill=blue!8] (input) {1.~Input: Route matching \& schema lookup};
    \node[layer, fill=orange!10, below=of input] (detect) {2.~Detection: Structural + Semantic};
    \node[layer, fill=green!8, below=of detect] (resolve) {3.~Resolution: DIRECT / CODEGEN};
    \node[layer, fill=red!8, below=of resolve] (safe) {4.~Safeguards: Validate $\rightarrow$ Ensemble $\rightarrow$ Fallback};
    \node[layer, fill=purple!8, below=of safe] (monitor) {5.~Monitoring: Latency, cost, accuracy};

    % Arrows between layers
    \draw[arrow] (input) -- (detect);
    \draw[arrow] (detect) -- (resolve);
    \draw[arrow] (resolve) -- (safe);
    \draw[arrow] (safe) -- (monitor);

    % Client and Backend
    \node[above=0.6cm of input, font=\scriptsize\itshape] (client) {Client (source schema)};
    \node[below=0.6cm of monitor, font=\scriptsize\itshape] (backend) {Backend (target schema)};
    \draw[arrow] (client) -- (input);
    \draw[arrow] (monitor) -- (backend);

    % LLM callout
    \node[llm, right=0.4cm of detect] (llm1) {LLM};
    \node[llm, right=0.4cm of resolve] (llm2) {LLM};
    \draw[-{Stealth[length=1.5mm]}, dashed] (detect.east) -- (llm1.west);
    \draw[-{Stealth[length=1.5mm]}, dashed] (resolve.east) -- (llm2.west);

    % Cache callout
    \node[llm, left=0.4cm of resolve] (cache) {Cache};
    \draw[-{Stealth[length=1.5mm]}, dashed] (resolve.west) -- (cache.east);
  \end{tikzpicture}
  \caption{SAGAI-MID five-layer pipeline architecture. Solid arrows show
    the data flow; dashed lines indicate LLM calls (layers~2--3) and
    cache lookups (layer~3). Layers~4--5 are fully deterministic.}
  \label{fig:architecture}
\end{figure}
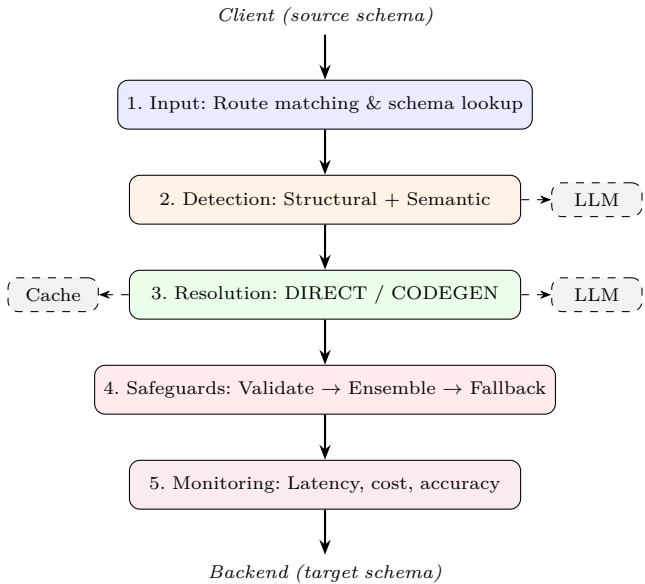

SAGAI-MID is an ASGI middleware that intercepts HTTP traffic between
clients and backend services, dynamically detecting and resolving
schema mismatches using LLMs. It is implemented as a five-layer pipeline
(Fig.~\ref{fig:architecture}) and integrates with existing service
architectures via FastAPI's \texttt{BaseHTTPMiddleware} pattern. The
middleware is transparent to both clients and backend services: clients
send requests in the source schema format, and the middleware transforms
them to the target schema before forwarding. We describe each layer
and then map the architecture to Bass~et~al.'s interoperability tactics.

\subsection{Input Layer}
\label{sec:arch:input}

The input layer intercepts HTTP requests with JSON bodies (POST, PUT,
PATCH). A \texttt{SchemaRegistry} maps routes to
\texttt{(source\_schema, target\_schema)} pairs via exact path matching
or longest-prefix matching. Each \texttt{RouteConfig} specifies source
and target JSON~Schemas, service identifiers, and the resolution
strategy to use. Non-registered routes pass through unmodified.

\subsection{Detection Module}
\label{sec:arch:detection}

The detection module uses a hybrid approach combining deterministic
structural analysis with LLM-powered semantic analysis.

\textbf{Structural Detector} (deterministic, no LLM call):
performs a recursive property walk comparing source and target
JSON~Schemas, detecting field missing/extra, type mismatches, nesting
mismatches, and cardinality mismatches. It classifies severity
(low/medium/high) based on type distance. This runs in under 1\,ms and
always executes first.

\textbf{Semantic Detector} (LLM-powered): sends both schemas to the
LLM with a structured output prompt, detecting naming mismatches
(e.g., \texttt{camelCase} $\leftrightarrow$ \texttt{snake\_case},
abbreviations) and unit mismatches (e.g., Celsius $\leftrightarrow$
Fahrenheit). It returns a \texttt{MismatchReport} via OpenAI structured
outputs (\texttt{.parse()}) with Pydantic response models. If the LLM
call fails, structural results remain valid (graceful degradation).

Semantic findings supersede structural findings for overlapping
(source\_path, target\_path) pairs, as they carry richer information.
The final merged \texttt{MismatchReport} is deduplicated.

\subsection{Resolution Engine}
\label{sec:arch:resolution}

The resolution engine implements two strategies, following
Falcao~et~al.\ \cite{falcao2026}, and embeds them in the middleware
pipeline.

\textbf{DIRECT Strategy}: a two-step LLM approach where {\em 1)}~the LLM
generates a \texttt{SchemaMapping} (field-level mappings with confidence
scores), and {\em 2)}~the LLM transforms the source data per-request using
the mapping. This is simple and handles arbitrary transformations, but
is non-deterministic across calls, slow (two LLM round-trips per
request), and expensive.

\textbf{CODEGEN Strategy}: a three-step approach where {\em 1)}~the LLM
generates a \texttt{SchemaMapping}, {\em 2)}~the LLM generates a Python
adapter function (\texttt{AdapterCode}), and {\em 3)}~the system compiles
and validates the generated code via \texttt{exec()}, caching the
compiled function. After the first invocation, subsequent requests
execute native Python with zero LLM calls.

Table~\ref{tab:strategies} summarizes the key differences.

\begin{table}[t]
  \caption{Comparison of Resolution Strategies}
  \label{tab:strategies}
  \centering
  \small
  \begin{tabular}{lcc}
    \toprule
    \textbf{Property} & \textbf{DIRECT} & \textbf{CODEGEN} \\
    \midrule
    LLM calls (cold) & 2/request & 3 (then 0) \\
    LLM calls (cached) & 1/request & 0 \\
    Deterministic & No & Yes (after compile) \\
    Cacheable & Mapping only & Mapping + code \\
    \bottomrule
  \end{tabular}
\end{table}

\textbf{Caching}: the mapping cache uses
\texttt{(SHA-256(source\_schema), SHA-256(target\_schema))} as key,
storing both \texttt{SchemaMapping} and compiled \texttt{AdapterCode}.
A cache hit for CODEGEN skips all LLM calls; for DIRECT, it skips the
mapping step but still requires per-request transformation.

\subsection{Safeguard Layer}
\label{sec:arch:safeguards}

The safeguard layer is a three-tier pipeline addressing LLM
non-determinism:

\textbf{Tier~1 -- Validation}: JSON~Schema validation (Draft~2020-12)
of the transformed output against the target schema, plus Pydantic model
validation of the LLM response structure and confidence threshold
checking. If valid, the result passes through without further safeguards.

\textbf{Tier~2 -- Ensemble Voting}: triggered when validation fails.
Runs $N$~parallel LLM calls (default $N\!=\!3$) for mapping generation,
taking majority vote on \texttt{(source\_field, target\_field)} pairs.
This follows self-consistency decoding~\cite{wang2023}, where sampling
multiple reasoning paths and taking majority vote improves accuracy,
and Parciak~et~al.'s finding that voting reduces hallucination in
schema matching~\cite{parciak2024}.

\textbf{Tier~3 - Rule-Based Fallback}: triggered when the ensemble
also fails. A fully deterministic layer with no LLM dependency:
\texttt{difflib} similarity matching for field renaming, a registry of
known unit conversions (C$\leftrightarrow$F, km/h$\leftrightarrow$mph,
meters$\leftrightarrow$feet), type coercion (string$\rightarrow$int,
ISO~8601$\rightarrow$epoch), and cardinality handling
(array$\leftrightarrow$single value). This tier never fails, guaranteeing
forward progress with best-effort output.

A key architectural decision is that safeguards are \emph{not} optional
add-ons but first-class pipeline stages with their own metrics
(ensemble trigger rate, fallback trigger rate, safeguard lift). This
distinguishes SAGAI-MID from prior work that evaluates LLM accuracy
without providing reliability mechanisms.

\subsection{Mapping to Interoperability Tactics}
\label{sec:arch:tactics}

Table~\ref{tab:tactics} maps SAGAI-MID's components to Bass~et~al.'s
interoperability tactics, showing how LLMs transform each tactic from a
static, design-time artifact into a dynamic, runtime capability.

\begin{table}[t]
  \caption{Mapping of SAGAI-MID to Interoperability Tactics~\cite{bass2021}}
  \label{tab:tactics}
  \centering
  \small
  \begin{tabular}{p{1.4cm}p{2.6cm}p{2.8cm}}
    \toprule
    \textbf{Tactic} & \textbf{Traditional} & \textbf{SAGAI-MID} \\
    \midrule
    Discover & Service registry (DNS, Consul) &
      Schema registry + LLM semantic matching \\
    \addlinespace
    Tailor Interface & Static adapter / decorator &
      LLM-generated translation layer \\
    \addlinespace
    Convert Data & Pre-built serializers, XSLT &
      LLM-generated conversion code \\
    \addlinespace
    Manage Resources & Connection pooling, rate limiting &
      Token budgets, model routing, semantic caching \\
    \bottomrule
  \end{tabular}
\end{table}

\subsection{Walkthrough: Weather API Migration}
\label{sec:arch:walkthrough}

To illustrate the pipeline, consider a weather service migrating from
REST~v1 (flat structure, Celsius, ISO~8601 timestamps) to v2 (nested
structure, Fahrenheit, Unix epochs). A client sends:

\begin{lstlisting}[numbers=none, basicstyle=\ttfamily\scriptsize]
{"city": "Amsterdam",
 "temperature_celsius": 18.5,
 "humidity_percent": 72,
 "wind_speed_kmh": 15.3,
 "timestamp": "2026-06-23T14:30:00Z"}
\end{lstlisting}

\textbf{Layer~1 (Input):} The \texttt{SchemaRegistry} matches the
route to a \texttt{RouteConfig} specifying source (v1) and target (v2)
JSON~Schemas.

\textbf{Layer~2 (Detection):} The structural detector identifies type
mismatches (\texttt{timestamp}: \texttt{string}$\rightarrow$\texttt{integer})
and nesting mismatches (\texttt{city}$\rightarrow$\texttt{location.name}).
The semantic detector identifies unit mismatches
(\texttt{temperature\_celsius}$\rightarrow$\texttt{temp\_f}: Celsius to
Fahrenheit) and naming mismatches
(\texttt{humidity\_percent}$\rightarrow$\texttt{humidity}).

\textbf{Layer~3 (Resolution):} Using the CODEGEN strategy, the LLM
generates a Python adapter function:

\begin{lstlisting}[numbers=none, basicstyle=\ttfamily\scriptsize]
def transform(data):
    return {
      "location": {"name": data["city"]},
      "measurements": {
        "temp_f": data["temperature_celsius"]
                  * 9/5 + 32,
        "humidity": float(
                    data["humidity_percent"]),
        "wind_mph": data["wind_speed_kmh"]
                    * 0.621371
      },
      "recorded_at": int(datetime.fromisoformat(
        data["timestamp"].replace("Z","+00:00")
      ).timestamp())
    }
\end{lstlisting}

This function is compiled via \texttt{exec()}, validated against sample
input, and cached under the
\texttt{(SHA-256(v1\_schema), SHA-256(v2\_schema))} key.

\textbf{Layer~4 (Safeguards):} The output is validated against the
v2 target schema using \texttt{jsonschema}. All fields are present
and correctly typed; Tier~1 passes, and no ensemble or fallback is
needed.

\textbf{Layer~5 (Monitoring):} Records latency, token count, and pass@1
result. Subsequent requests execute the cached function with zero LLM
calls ($<$1\,ms). The output matches the v2 target schema:

\begin{lstlisting}[numbers=none, basicstyle=\ttfamily\scriptsize]
{"location": {"name": "Amsterdam"},
 "measurements": {"temp_f": 65.3,
   "humidity": 72.0, "wind_mph": 9.51},
 "recorded_at": 1782225000}
\end{lstlisting}

% ══════════════════════════════════════════════════════════════════════
% IV. IMPLEMENTATION
% ══════════════════════════════════════════════════════════════════════

\section{Implementation}
\label{sec:implementation}

\subsection{Technology Stack}
\label{sec:impl:stack}

SAGAI-MID is implemented in Python~3.13 using FastAPI with Uvicorn for
async-native ASGI middleware support. The middleware extends
\texttt{BaseHTTPMiddleware}, which intercepts every HTTP request and
response. For registered routes, the middleware reads the request body,
runs the detect$\rightarrow$resolve$\rightarrow$safeguard pipeline, and
replaces the body before forwarding; for unregistered routes, the
request passes through unmodified with zero overhead.

LLM calls use the OpenAI API (compatible with both OpenAI and xAI
endpoints) with structured outputs~\cite{openai2024} - the
\texttt{.parse()} method with Pydantic response models guarantees
valid JSON conforming to the expected schema, functioning as runtime
contract enforcement~\cite{meyer1992}. Five Pydantic models define the
LLM response contracts: \texttt{MismatchReport}, \texttt{SchemaMapping},
\texttt{FieldMapping}, \texttt{AdapterCode}, and
\texttt{TransformedData}. Validation uses both Pydantic~v2 (LLM
response structure) and \texttt{jsonschema} Draft~2020-12 (target
schema conformance). All LLM calls are asynchronous (\texttt{await}
on \texttt{AsyncOpenAI}).

\subsection{Prompt Engineering}
\label{sec:impl:prompts}

Four prompt files are stored as plain text files in a dedicated
\texttt{prompts/} directory, editable without code changes:
{\em 1)}~\texttt{detect\_mismatch.txt} instructs the LLM to compare source
and target schemas and report mismatches with type classification;
{\em 2)}~\texttt{generate\_mapping.txt} requests field-level mapping with
confidence scores and transformation expressions;
{\em 3)}~\texttt{generate\_adapter.txt} asks for a complete Python function
that transforms source data to match the target schema, including
import statements and type conversions; and
{\em 4)}~\texttt{transform\_data.txt} provides the mapping and source data
for per-request transformation.

Structured outputs guarantee valid JSON conforming to the expected
Pydantic model, so prompts contain no output format instructions, only
the task description and schema context~\cite{openai2024}. This
separation simplifies prompt engineering~\cite{wei2022} and reduces
token consumption. For non-reasoning models, we set
temperature\,=\,0.0 for maximum determinism. Reasoning models (GPT-5,
GPT-5-nano, Grok~4.1~Fast~R) do not accept the \texttt{temperature}
parameter and require longer timeouts.

\subsection{Model Adaptation Layer}
\label{sec:impl:models}

The LLM client adapts to behavioral differences across model families.
GPT-5 reasoning models (GPT-5, GPT-5-nano) use
\texttt{reasoning\_effort} instead of \texttt{temperature}; Grok
reasoning models (Grok~4.1~Fast~R) accept neither parameter. GPT-5.2
supports \texttt{temperature} only with
\texttt{reasoning\_effort="none"}. All reasoning models require
120+\,s timeouts (vs.\ 60\,s for standard models). This adaptation is
transparent to the rest of the pipeline.

\subsection{Evaluation Scenarios}
\label{sec:impl:scenarios}

We design 10~interoperability scenarios covering diverse mismatch types
across three protocols (Table~\ref{tab:scenarios}). Each scenario is a
self-contained JSON fixture with source/target JSON~Schemas, sample
input data, golden reference output, and expected mismatches for
detection evaluation. Three mock services, Weather REST (v1/v2),
IoT~Sensor REST bridge, and Stock GraphQL (Strawberry), cover REST,
GraphQL, and IoT/MQTT protocols.

\begin{table}[t]
  \caption{Evaluation Scenarios and Mismatch Types}
  \label{tab:scenarios}
  \centering
  \small
  \begin{tabular}{clp{2.6cm}}
    \toprule
    \textbf{\#} & \textbf{Scenario} & \textbf{Mismatch Types} \\
    \midrule
    1 & Weather v1$\rightarrow$v2 & Nesting, unit, name, type \\
    2 & IoT sensor$\rightarrow$analytics & Unit (C$\rightarrow$F), name \\
    3 & Stock REST$\rightarrow$GraphQL & Name (camelCase) \\
    4 & Multi-sensor aggregation & Cardinality, aggregate \\
    5 & Date format bridging & Type (ISO$\rightarrow$epoch) \\
    6 & Nested$\rightarrow$flat device log & Nesting (flatten) \\
    7 & Metric normalization & Name (OpenTelemetry) \\
    8 & Missing optional fields & Field missing/extra \\
    9 & Array$\leftrightarrow$single value & Cardinality \\
    10 & Combined complex & All types combined \\
    \bottomrule
  \end{tabular}
\end{table}

% ══════════════════════════════════════════════════════════════════════
% V. EVALUATION
% ══════════════════════════════════════════════════════════════════════

\section{Evaluation}
\label{sec:evaluation}

\subsection{Experimental Setup}
\label{sec:eval:setup}

We evaluate all 10~scenarios with both DIRECT and CODEGEN strategies,
with and without safeguards (4~combinations), for 3~runs per
combination (40~scenario-strategy-mode runs per model). We test
six~LLMs from two providers: GPT-4o, GPT-5, GPT-5.2, and GPT-5-nano
from OpenAI; and Grok~4.1~Fast in both non-reasoning and reasoning
modes from xAI. The mapping cache is disabled during benchmarking to
measure raw LLM performance. Each run records per-stage latency, token
consumption, and the complete transformed output.

\subsection{Metrics}
\label{sec:eval:metrics}

\textbf{Correctness.} \emph{pass@1}~\cite{chen2021}: binary exact match
against the golden reference (using DeepDiff~\cite{deepdiff} with
float tolerance $\epsilon\!=\!0.01$ and numeric type equivalence).
\emph{Field~F1}: leaf-key overlap between actual and expected outputs.
\emph{Value accuracy}: fraction of shared keys where values match.
\emph{Detection precision/recall}: mismatch pair overlap vs.\ expected.

\textbf{Performance.} End-to-end latency (ms) across detect, resolve,
and safeguard stages. We report mean and P95 across runs.

\textbf{Cost.} Total token consumption and estimated USD cost per model
(token count $\times$ per-model pricing) across all 40~combinations.

\subsection{Results}
\label{sec:eval:results}

Table~\ref{tab:crossmodel} presents the main cross-model results.
Across all six~models, CODEGEN achieves equal or higher mean pass@1
than DIRECT. The best-performing configuration, Grok~4.1~Fast
(reasoning) with CODEGEN, achieves 0.90~pass@1 at the lowest total
cost (\$0.18). Field~F1 is universally $\geq\!0.98$, indicating that
all models correctly identify which fields to map; the differentiator
is \emph{value accuracy}, whether transformed values are correct.

\begin{table*}[t]
  \caption{Cross-Model Results (Mean over 10 Scenarios, 3~Runs per Combination, 40~Combinations per Model)}
  \label{tab:crossmodel}
  \centering
  \small
  \begin{tabular}{llcccccccc}
    \toprule
    & & \multicolumn{2}{c}{\textbf{pass@1}} & \multicolumn{2}{c}{\textbf{Value Acc.}} & \multicolumn{2}{c}{\textbf{Mean Latency (s)}} & \\
    \cmidrule(lr){3-4} \cmidrule(lr){5-6} \cmidrule(lr){7-8}
    \textbf{Model} & \textbf{Provider} & \textbf{D} & \textbf{C} & \textbf{D} & \textbf{C} & \textbf{D} & \textbf{C} & \textbf{Total Cost} \\
    \midrule
    GPT-4o & OpenAI & 0.70 & 0.83 & 0.92 & 0.97 & 11.8 & 16.1 & \$2.43 \\
    GPT-5 & OpenAI & 0.80 & 0.80 & 0.95 & 0.95 & 75.4 & 103.8 & \$6.25 \\
    GPT-5.2 & OpenAI & 0.83 & 0.87 & 0.97 & 0.97 & 31.7 & 36.7 & \$4.85 \\
    GPT-5-nano & OpenAI & 0.73 & 0.73 & 0.95 & \textbf{0.98} & 50.6 & 55.1 & \$0.27 \\
    Grok 4.1 Fast (NR) & xAI & 0.70 & 0.87 & 0.92 & \textbf{0.98} & \textbf{9.2} & 10.5 & \$0.19 \\
    Grok 4.1 Fast (R) & xAI & 0.87 & \textbf{0.90} & 0.97 & \textbf{0.98} & 69.4 & 70.8 & \textbf{\$0.18} \\
    \midrule
    \textbf{Mean} & & 0.77 & 0.83 & 0.95 & 0.97 & 41.3 & 48.8 & \$2.36 \\
    \bottomrule
  \end{tabular}
  \begin{flushleft}
    \footnotesize D\,=\,DIRECT, C\,=\,CODEGEN, NR\,=\,non-reasoning, R\,=\,reasoning.
    Total cost covers all 40~scenario-strategy-mode combinations $\times$ 3~runs.
    Bold indicates best value per column.
  \end{flushleft}
\end{table*}

Table~\ref{tab:perscenario} shows per-scenario pass@1 averaged across
all models, revealing the difficulty spectrum. Four scenarios
(stock casing, nested-to-flat, missing fields, array-single) achieve
perfect 1.00~pass@1 across all models. Sensor analytics (unit
conversion) and metric normalization (OpenTelemetry renaming) are the
hardest, achieving only 0.50--0.56 mean pass@1, with significant
model-dependent variation.

\begin{table}[t]
  \caption{Per-Scenario pass@1 (Mean across All 6~Models)}
  \label{tab:perscenario}
  \centering
  \small
  \begin{tabular}{clcc}
    \toprule
    \textbf{\#} & \textbf{Scenario} & \textbf{D} & \textbf{C} \\
    \midrule
    1 & Weather version & 0.56 & 0.78 \\
    2 & Sensor analytics & 0.56 & 0.50 \\
    3 & Stock casing & 1.00 & 1.00 \\
    4 & Multi-sensor & 1.00 & 0.95 \\
    5 & Date bridging & 0.56 & 0.95 \\
    6 & Nested to flat & 1.00 & 1.00 \\
    7 & Metric normalization & 0.50 & 0.56 \\
    8 & Missing fields & 1.00 & 1.00 \\
    9 & Array single & 1.00 & 1.00 \\
    10 & Combined complex & 0.56 & 0.61 \\
    \midrule
    \textbf{Mean} & & \textbf{0.77} & \textbf{0.84} \\
    \bottomrule
  \end{tabular}
\end{table}

Table~\ref{tab:performance} presents latency and cost metrics. Latency
varies by over 10$\times$ across models: Grok~4.1~Fast (non-reasoning)
is the fastest at 9--11\,s mean, while GPT-5 is the slowest at
75--104\,s. Cost varies by over 30$\times$: from \$0.18 (Grok~4.1~Fast
reasoning) to \$6.25 (GPT-5). Notably, the most accurate model
(Grok~4.1~R, 0.90~pass@1) is also the cheapest, while the most
expensive model (GPT-5, \$6.25) achieves only 0.80~pass@1.

\begin{table}[t]
  \caption{Performance and Cost Comparison}
  \label{tab:performance}
  \centering
  \small
  \begin{tabular}{lccc}
    \toprule
    \textbf{Model} & \textbf{P95 Lat.} & \textbf{Tokens} & \textbf{Cost} \\
    & \textbf{(s)} & \textbf{(total)} & \textbf{(\$)} \\
    \midrule
    GPT-4o & 14--19 & 615K & 2.43 \\
    GPT-5 & 88--117 & 1,180K & 6.25 \\
    GPT-5.2 & 36--40 & 943K & 4.85 \\
    GPT-5-nano & 57--64 & 1,317K & 0.27 \\
    Grok 4.1 (NR) & 10--11 & 721K & 0.19 \\
    Grok 4.1 (R) & 80--81 & 1,535K & 0.18 \\
    \bottomrule
  \end{tabular}
  \begin{flushleft}
    \footnotesize P95~Latency range shows DIRECT--CODEGEN.
    Tokens and cost are totals across all 40~combinations $\times$ 3~runs.
  \end{flushleft}
\end{table}

\subsection{Analysis}
\label{sec:eval:analysis}

\textbf{CODEGEN outperforms DIRECT.}
Across all six models, CODEGEN achieves equal or higher mean~pass@1
than DIRECT (0.83 vs.\ 0.77 overall). The advantage is most pronounced
on complex scenarios: for weather version migration (scenario~1),
CODEGEN achieves 0.78 vs.\ 0.56 DIRECT; for date bridging (scenario~5),
0.95 vs.\ 0.56. The CODEGEN strategy benefits from code validation---the
generated adapter function is tested against sample input before
caching, catching errors that would propagate silently in DIRECT mode.

\textbf{Model-dependent difficulty patterns.}
We observe a striking pattern: models exhibit complementary strengths
across scenario types. Reasoning models (GPT-5, Grok~4.1~R) excel at
complex combined scenarios (scenario~10: both achieve 1.00~pass@1) but
struggle with sensor analytics unit conversion (scenario~2: both
achieve 0.00~pass@1). Conversely, non-reasoning models (GPT-4o,
Grok~4.1~NR) handle unit conversions well (1.00~DIRECT pass@1 on scenario~2)
but struggle with the combined scenario (GPT-4o achieves 0.00~pass@1
on scenario~10 with CODEGEN).

Similarly, metric normalization (scenario~7) is consistently solved by
GPT-4o and both Grok models (all 1.00~pass@1) but fails across all
three OpenAI reasoning models: GPT-5 (0.00), GPT-5.2 (0.00--0.33),
and GPT-5-nano (0.00). This suggests that reasoning models may
overthink straightforward mapping tasks that require simple pattern
matching rather than deep reasoning.

\textbf{Cost-accuracy frontier.}
The relationship between cost and accuracy is non-monotonic
(Table~\ref{tab:frontier}). The cheapest model (Grok~4.1~R at \$0.18)
achieves the \emph{highest} accuracy (0.90~pass@1~CODEGEN), while the
most expensive model (GPT-5 at \$6.25) achieves only 0.80. This
finding challenges the assumption that larger, more expensive models
necessarily produce better results for structured transformation tasks.
The best cost-performance ratio for latency-sensitive applications is
Grok~4.1~Fast~(NR) at 10\,s mean latency, 0.87~CODEGEN~pass@1, and
\$0.19~total.

\begin{table}[t]
  \caption{Cost-Accuracy-Latency Comparison (CODEGEN Strategy)}
  \label{tab:frontier}
  \centering
  \small
  \begin{tabular}{@{}l@{\hspace{2pt}}c@{\hspace{6pt}}c@{\hspace{6pt}}c@{}}
    \toprule
    \textbf{Model} & \textbf{pass@1} & \textbf{Latency} & \textbf{Cost} \\
    \midrule
    Grok R & \textbf{0.90} & 70.8\,s & \textbf{\$0.18} \\
    Grok NR & 0.87 & \textbf{10.5\,s} & \$0.19 \\
    GPT-5.2 & 0.87 & 36.7\,s & \$4.85 \\
    GPT-4o & 0.83 & 16.1\,s & \$2.43 \\
    GPT-5 & 0.80 & 103.8\,s & \$6.25 \\
    GPT-5-nano & 0.73 & 55.1\,s & \$0.27 \\
    \bottomrule
  \end{tabular}
  \begin{flushleft}
    \footnotesize Bold indicates best value per column.
  \end{flushleft}
\end{table}

\textbf{Safeguard impact.}
The safeguard layer shows modest aggregate impact on pass@1 across
models that already achieve high baseline accuracy. Ensemble voting
activates on 0--100\% of runs depending on model and scenario. For
GPT-5.2, safeguards provide measurable improvement on specific
scenarios (sensor analytics DIRECT: +0.33, metric normalization
CODEGEN: +0.33). However, we also observe cases where safeguards
marginally reduce accuracy (e.g., GPT-5 sensor analytics CODEGEN:
$-$0.33), likely due to ensemble voting selecting a less accurate
consensus mapping.

The primary value of the safeguard architecture is as a \emph{safety
net} rather than an accuracy booster: it prevents catastrophic failures
by providing deterministic fallback when LLM outputs fail validation,
ensuring forward progress on every request. The rule-based fallback
(Tier~3) handles known unit conversions and type coercions
deterministically, providing a guaranteed minimum quality floor.

\textbf{Detection quality.}
Detection recall is consistently high across all models ($\geq$0.83),
meaning the system identifies the vast majority of schema mismatches.
Detection precision varies more widely (0.46--1.00). GPT-4o achieves
the highest detection precision (0.95) with perfect recall (1.00),
while reasoning models tend to over-report mismatches (lower precision).
Importantly, lower detection precision does not impair resolution
quality; extra reported mismatches are benign, as the resolution
engine only acts on relevant ones.

\subsection{Threats to Validity}
\label{sec:eval:threats}

\textbf{Internal}: LLM outputs exhibit non-determinism even at
temperature\,=\,0 (documented OpenAI behavior). Golden outputs are
hand-crafted, introducing potential human error. With $N\!=\!3$ runs per
combination, variance estimates have wide confidence intervals.

\textbf{External}: 10~scenarios may not cover all interoperability
patterns. We evaluate JSON-only payloads (no XML, Protobuf, Avro) over
HTTP-only transport. A single middleware deployment (not distributed or
multi-hop) is tested.

\textbf{Construct}: pass@1 is a strict binary metric that may
undercount near-correct outputs. Token cost depends on pricing at time
of evaluation (February 2026), which may change.

% ══════════════════════════════════════════════════════════════════════
% VI. DISCUSSION
% ══════════════════════════════════════════════════════════════════════

\section{Discussion}
\label{sec:discussion}

\textbf{LLMs as runtime architectural components.}
Our results demonstrate that LLMs can effectively serve as runtime
interoperability components, achieving 0.73--0.90~pass@1 on diverse
schema transformation tasks. This enables a paradigm shift: the
interoperability tactics from Bass~et~al., traditionally implemented
as static, design-time adapters, become dynamic runtime capabilities.
The practical implication is that new service integrations can be
handled without manual adapter development or deployment cycles, at
the cost of LLM latency and API expenses. This is particularly
relevant in microservice ecosystems where Lercher~et~al.\
\cite{lercher2024} report that API evolution is a persistent
maintenance burden.

\textbf{Model selection guidance.}
Our cross-model evaluation reveals that model selection for structured
transformation tasks differs from general-purpose LLM benchmarks. The
most expensive model (GPT-5 at \$6.25) is not the most accurate
(0.80~pass@1), while the cheapest (Grok~4.1~R at \$0.18) achieves the
highest accuracy (0.90~pass@1). We recommend: (1)~Grok~4.1~Fast~(NR)
for latency-critical applications (10\,s mean, 0.87~pass@1);
(2)~Grok~4.1~Fast~(R) for accuracy-critical applications (0.90~pass@1
at minimal cost); and (3)~GPT-4o as a balanced option with the best
detection precision.

\textbf{When to use DIRECT vs.\ CODEGEN.}
CODEGEN should be preferred for production workloads: it achieves
higher pass@1, is deterministic after first invocation, and amortizes
LLM cost over repeated requests. DIRECT is appropriate for prototyping,
one-off transformations, or scenarios where schema pairs change
frequently. A hybrid approach (starting with DIRECT for new schema pairs,
graduating to CODEGEN after stabilization) optimizes for both agility
and reliability. The CODEGEN advantage is most pronounced on complex
scenarios involving multiple mismatch types simultaneously (e.g.,
weather version migration: 0.78 vs.\ 0.56 pass@1).

\textbf{Safeguard architecture as defense-in-depth.}
While safeguards show modest aggregate impact on pass@1, their value is
architectural: they provide defense-in-depth against LLM
non-determinism. The three-tier design ensures that every request
receives a response, degrading gracefully from LLM-generated to
ensemble-voted to rule-based output. This is critical for production
middleware where a failed transformation is worse than a best-effort one.
The safeguard architecture directly addresses the SAGAI workshop's
topic of ``tactics to deal with the non-deterministic character of
generative~AI.''

\textbf{Quality attribute tradeoffs.}
Introducing LLMs as runtime architectural components creates
fundamental tradeoffs across quality attributes. We analyze these
through the lens of ISO/IEC~25010~\cite{iso25010}:

\emph{Performance efficiency vs.\ modifiability}: Static adapters
execute in $<$1\,ms but require manual coding for every schema change.
SAGAI-MID incurs 10--104\,s latency on cold starts but handles novel
schema combinations without code changes; CODEGEN caching converges
toward static adapter performance on subsequent requests.

\emph{Reliability vs.\ flexibility}: LLM non-determinism introduces a
new failure mode. The three-tier safeguard architecture provides
progressively more deterministic fallbacks: ensemble voting reduces
variance~\cite{wang2023}, and rule-based fallback eliminates
non-determinism entirely.

\emph{Cost vs.\ autonomy}: LLM API costs (\$0.18--\$6.25) are a new
operational expense, but replace developer time for manual adapter
development. CODEGEN caching amortizes costs, and model selection can
reduce them by over 30$\times$.

\textbf{Reasoning model paradox.}
Reasoning models (GPT-5, GPT-5-nano, Grok~4.1~R) struggle with
seemingly simple tasks: unit conversions (scenario~2) and metric
normalization (scenario~7) yield 0.00~pass@1 where non-reasoning
models achieve 1.00. Conversely, reasoning models excel at complex
combined transformations (scenario~10). This parallels Liu~et~al.'s
\cite{liu2025} finding that chain-of-thought prompting can \emph{reduce}
performance on tasks where deliberation hurts human accuracy---extended
inference chains introduce error accumulation on mechanistic tasks
(e.g., $F = C \times 9/5 + 32$). This suggests model routing based on
mismatch complexity could further improve system-level accuracy.

\textbf{Limitations.}
Key limitations include: (1)~LLM latency (10--104\,s on cold starts)
compared to static adapters ($<$1\,ms), making this unsuitable for
ultra-low-latency requirements without caching; (2)~cloud API
dependency for LLM calls (cost, network availability, vendor lock-in);
(3)~security considerations around \texttt{exec()} of LLM-generated
code in CODEGEN, which requires sandboxing in production;
(4)~JSON-only payload support (no XML, Protobuf, or binary formats);
(5)~prompt sensitivity, performance depends on prompt engineering
quality and model-specific tuning; and (6)~evaluation scale, 10
scenarios with 3~runs each provide limited statistical power for
per-scenario claims.

% ══════════════════════════════════════════════════════════════════════
% VII. CONCLUSION & FUTURE WORK
% ══════════════════════════════════════════════════════════════════════

\section{Conclusion}
\label{sec:conclusion}

We presented SAGAI-MID, a generative AI-driven middleware that uses
LLMs to dynamically detect and resolve schema mismatches at runtime.
The five-layer architecture connects LLM-based middleware to established
interoperability tactics from Bass~et~al., transforming them from
static design-time artifacts into dynamic runtime capabilities. A
hybrid detection module combines deterministic structural analysis with
LLM semantic analysis for comprehensive mismatch identification. Dual
resolution strategies (DIRECT and CODEGEN) offer
flexibility-determinism tradeoffs with hash-keyed caching, and a
three-tier safeguard stack provides defense-in-depth against LLM
non-determinism.

Our evaluation across 10~interoperability scenarios (spanning REST
version migration, IoT-to-analytics bridging, and GraphQL protocol
conversion) and six~LLMs from two providers reveals four key findings:
(1)~CODEGEN consistently outperforms DIRECT (0.83 vs.\ 0.77
mean~pass@1), with the advantage most pronounced on complex
multi-mismatch scenarios; (2)~the best model achieves 0.90~pass@1 at
the lowest cost (\$0.18), demonstrating a non-monotonic
cost-accuracy relationship; (3)~reasoning and non-reasoning models
exhibit complementary strengths, suggesting that model routing could
further improve accuracy; and (4)~safeguards provide architectural
value as a defense-in-depth mechanism rather than a consistent
accuracy booster, with the rule-based fallback guaranteeing forward
progress on every request.

These findings have practical implications for software architects:
LLMs can serve as runtime interoperability components for systems
where schema evolution outpaces manual adapter development, provided
that appropriate safeguards, caching, and model selection strategies
are in place.

Future work includes: (1)~local/on-premise models to eliminate cloud
dependency and address data privacy, Falcao~et~al.\ show open-source
models achieving competitive results~\cite{falcao2026};
(2)~model routing based on mismatch complexity to exploit the
complementary strengths of reasoning and non-reasoning models;
(3)~service mesh integration (e.g., as an Envoy or Istio filter) for
transparent deployment in microservice architectures;
(4)~schema evolution tracking for proactive adapter regeneration when
backend schemas change; and (5)~broader protocol support including
XML, Protobuf, gRPC, and binary formats.

% ══════════════════════════════════════════════════════════════════════
% REFERENCES
% ══════════════════════════════════════════════════════════════════════

\bibliographystyle{IEEEtran}

\balance

\end{document}